\documentclass[twocolumn,prl,showpacs,superscriptaddress]{revtex4}
\usepackage{graphicx}
\usepackage{amsmath}
\usepackage{amssymb}
\begin{document}

\title{Polaronic state and nanometer-scale phase separation in colossal
magnetoresistive manganites}

\author{Sahana~R\"o\ss ler}
\affiliation{Max Planck Institute for Chemical Physics of Solids,
N\"othnitzer Stra\ss e~40, 01187 Dresden, Germany}
\author{S.~Ernst}
\affiliation{Max Planck Institute for Chemical Physics of Solids,
N\"othnitzer Stra\ss e~40, 01187 Dresden, Germany}
\author{B.~Padmanabhan}
\affiliation{Department of Physics, Indian Institute of Science,
Bangalore 560012, India}
\author{Suja~Elizabeth}
\affiliation{Department of Physics, Indian Institute of Science,
Bangalore 560012, India}
\author{H.~L.~Bhat}
\affiliation{Department of Physics, Indian Institute of Science,
Bangalore 560012, India}
\author{F.~Steglich}
\affiliation{Max Planck Institute for Chemical Physics of Solids,
N\"othnitzer Stra\ss e~40, 01187 Dresden, Germany}
\author{S.~Wirth}
\affiliation{Max Planck Institute for Chemical Physics of Solids,
N\"othnitzer Stra\ss e~40, 01187 Dresden, Germany}
\date{\today}

\begin{abstract}
High resolution topographic images obtained by scanning tunneling
microscope in the insulating state of
Pr$_{0.68}$Pb$_{0.32}$MnO$_3$ single crystals showed regular
stripe-like or zigzag patterns on a width scale of 0.4--0.5 nm
confirming a high temperature polaronic state. Spectroscopic
studies revealed inhomogeneous maps of zero-bias conductance with
small patches of metallic clusters on length scale of 2--3 nm
only within a narrow temperature range close to the
metal-insulator transition. The results give a direct observation
of polarons in the insulating state, phase separation of
nanometer-scale metallic clusters in the paramagnetic metallic
state, and a homogeneous ferromagnetic state.
\end{abstract}
 \maketitle

There is an intense research going on to understand the
remarkable and complex properties shown by a whole family of
strongly correlated electron systems.  In these materials, a
self-inflicted, spontaneous instability of the electronic state
and competing long-range interactions may result in the formation
of nanometer-sized regions of different phases. Such states have
been considered as charge and spin ordered stripes in under-doped
cuprates \cite{kiv}, as polar domains in relaxor ferroelectrics
\cite{find}, or as phase separation (PS) between insulating
paramagnetic (pm) and conducting ferromagnetic regions \cite{dag}
in mixed valence manganites of perovskite type $A$MnO$_3$
($A$--rare earth or doped divalent ion).
%The various unconventional macroscopic properties of under-doped
%cuprates, slow dynamics in relaxor ferroelectrics or the colossal
%magnetoresistance (CMR) in manganites have been associated with the
%evolution of nanometer-scale electronic inhomogeneities or ordering.
In the latter case,
%ABO$_3$ (A--rare earth/divalent ion, B--Mn$^{3+}$/Mn$^{4+}$ ion)
evidence for PS has been found by various experimental
techniques, such as electron microscopy \cite{ueh}, scanning
tunneling microscopy/spectroscopy (STM/S) \cite{fat,bec},
magnetic force microscopy \cite{zha} and photoelectron
spectroscopy \cite{sar}. These experiments showed inhomogeneities
of random shape on a length scale of several hundred nanometers.
Further, the PS persisted deep into the metallic state in some of
these manganites. However, computational studies on models of
manganites considering double exchange, Jahn-Teller (JT)
interaction and long range Coulomb potential could show only
regularly spaced nanometer-scale PS \cite{mal}. The random
location and shape of the clusters observed experimentally
\cite{ueh,fat,bec,zha,sar} are conjectured to be caused by
quenched disorder in the couplings induced by chemical
substitution \cite{mo1,mo2}. Recent STS studies combined with
transmission electron microscopy \cite{mos} on $A$-site ordered
and disordered La$_{0.75}$Ca$_{0.25}$MnO$_3$ thin films showed
that PS persists in the metallic state only in the disordered
film. But, this study does not address the question of PS at the
metal-insulator transition temperature, $T_{MI}$. Thus, the
origin of the PS, the length scale involved, the role of quenched
disorder originating from the random $A$-site substitution, and
the temperature range at which PS occurs, remain all strongly
debated.

The polaron effect due to strong JT electron-phonon coupling is
considered central to understand the remarkable transport
properties, specifically the colossal magnetoresistance (CMR), of
manganites \cite{mil}. The high temperature polaronic state in
reciprocal space has been probed experimentally \cite{vas,nel}
revealing complex polaron effects such as polaron correlation,
polaron ordering, and charge localization. Using STM, polarons can
be imaged directly in the real space. Polaron confinement was
recently observed in a layered manganite single crystal using STM
\cite{ron}. Charge ordering was reported for thin films
(La$_{5/8-x}$Pr$_x$)Ca$_{3/8}$MnO$_3$ [short range charge exchange
(CE) type] in the pm state \cite{ma} and for highly doped
Bi$_{0.24}$Ca$_{0.76}$MnO$_3$ single crystals \cite{ren-nat}.

Spatially resolved STS measurements \cite{fat} on thin films of
La$_{0.73}$Ca$_{0.27}$MnO$_3$ on SrTiO$_3$ showed coexistence of
regions with metallic, insulating as well as intermediate
conductivities, extending over several hundred nanometers.
However, these images were obtained at a rather high fixed bias
voltage of 3 V (much larger than the semiconducting gap of
0.2--0.3 V in manganites) and may not reflect the ground state
properties. On the other hand, in Ref.~\onlinecite{bec} the
zero-bias conductance, $G_0 = dI/dV|_{V=0}$, of
La$_{0.7}$Sr$_{0.3}$MnO$_3$/MgO thin films was mapped as a
function of temperature $T$, and a threshold criterion was applied
to distinguish metallic and insulating regions. Such a threshold
criterion will not give an unambiguous evidence for the existence
of PS because any statistical distribution of conductance, whose
average value shifts with $T$, will seem to show PS \cite{ren}.
Further, in thin film samples, miss-fit strain induced by the
substrates seems to influence the electrical properties
\cite{par}.

To resolve some of these issues from experimental side, we
carried out STM/S on Pr$_{0.68}$Pb$_{0.32}$MnO$_3$ (PPMO) single
crystals providing largely strain-free materials. We address two
important questions in the physics of manganites, namely, the high
temperature polaronic state and the nanometer-scale electronic PS.
We present STM images of polaronic Mn$^{3+}$-sites and doped hole
localization on Mn$^{4+}$-sites with atomic resolution. In
addition, we provide clear evidence for nanometer-scale PS and
percolation just below $T_{MI}$ by looking at the entire
distribution of $G_0$ and its dependence on $T$. Information on
the length scale of the inhomogeneities and the $T$ range within
which it appears is obtained. We discuss the role of quenched
disorder or doping and compare the results with macroscopic
properties of the same single crystal.

Single crystals of PPMO used for the present study were taken from
a batch of crystals, whose preparation and properties were already
\begin{figure}[tb]
\centering \includegraphics[width=7.8cm,clip]{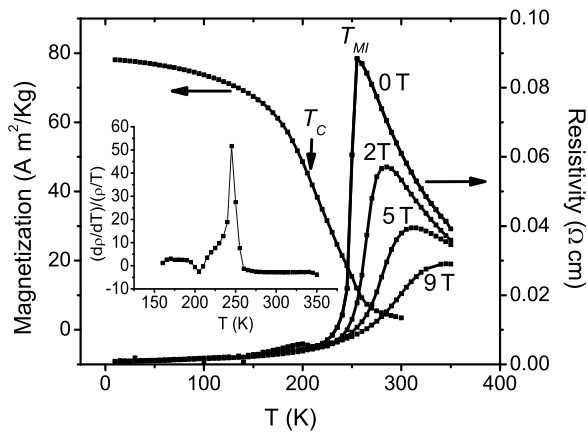}
\caption{Temperature dependence of magnetization (left scale) and
resistivity (right scale) of Pr$_{0.68}$Pb$_{0.32}$MnO$_3$ single
crystals measured at different magnetic fields. Inset: The
logarithmic derivative $(d\rho /dT)/(\rho /T)$ as function of $T$
at $H =0$.} \label{mag}
\end{figure}
reported in \cite{pa1,pa2}. In Pr$_{1-x}$Pb$_x$MnO$_3$, the Curie
temperature $T_C$ and $T_{MI}$ do not coincide, and metal-like
conductivity occurs in a pm state in parts of the phase diagram
\cite{pa1,li}, a phenomenon uncommon to mixed valence manganites.
Fig.~\ref{mag} shows the temperature dependence of magnetization
($M$), resistivity ($\rho$) and magnetoresistive properties of a
PPMO sample. The magnetoresistance, $[\rho(H) - \rho(0)] /
\rho(0)$, is found to be $\sim$90\% close to $T_{MI}$ under a
field of 9 T. From the maximum change in slope of the $M$ {\it
vs.} $T$ curve, $T_C \approx$ 210 K was estimated which is about
45 K lower than the corresponding $T_{MI}\approx$ 255 K. Such an
approach to estimate $T_C$ is supported by elaborate
investigations on a similar Pr$_{0.7}$Pb$_{0.3}$MnO$_3$ single
crystal in which the so-determined $T_C \sim$ 197 K agrees well
with results from detailed static magnetization scaling analysis
\cite{pa2} as well as heat capacity measurements \cite{pa3}
($T_{MI} \approx$ 235 K in this compound \cite{pa1}). The scaling
analysis embracing the critical temperature indicated that the
underlying magnetic transition is a conventional one, with
short-range Heisenberg-like critical exponents. This study
emphasizes on the presence of additional frustrated couplings
which intercepts the formation of long range order. Deviation of
the inverse susceptibility from the Curie-Weiss law above $T_C$
\cite{pa1} and history-dependent transport properties \cite{li}
suggest a
\begin{figure}[tb]
\centering \includegraphics[width=8.0cm,clip]{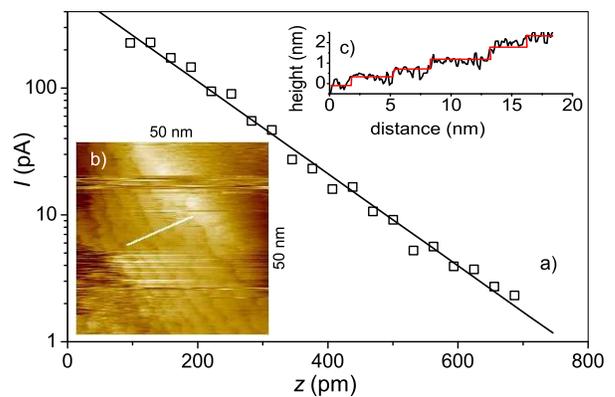} \caption{(a)
Dependence of tunneling current $I$ on relative tip-sample
distance $z$ on a semi-logarithmic scale. (b) Surface topography
over an area 50$\,\times\,$50 nm$^2$. (c) Height profile along the
white line drawn in inset (b).} \label{Iz}
\end{figure}
presence of small magnetic metallic clusters above $T_C$ that form
percolating metallic paths upon reducing $T$ in the pm metallic
state. Note that evidence for the formation of localized
$\sim$1.2 nm magnetic clusters above $T_C$ in another mixed
valent manganite has earlier been found by small-angle neutron
scattering measurements \cite{det}. We also note the sharpness of
the resistance transition which can be inferred from the
logarithmic derivative of the resistance $(d\rho / dT) / (\rho /
T)$ plotted in the inset of Fig.~\ref{mag}. Such a sharp
metal-insulator transition is indicative of a strain-free sample
\cite{par} (the tolerance factor, $t =$ 0.965, is close to unity
indicating good ionic size match).

For the tunneling studies a STM (Omicron Nanotechnology) under
ultra high vacuum conditions ($p \le 10^{-10}$ mbar) was utilized
at eleven fixed temperatures, 30 K $\le T \le$ 300 K, mostly in
the vicinity of $T_C$ and $T_{MI}$. Since crystals with perovskite
structure do not cleave easily, preparation of a clean surface
for the STM is a challenge. Just before inserting the crystal
into the UHV chamber, we thoroughly cleaned the crystal surface
in isopropanol using an ultrasonic bath and then, inside
isopropanol, scraped the surface to rip off some part of the
surface. This preparation gave us clean surfaces on a length
scale of microns. STM was conducted using tungsten tips, and
typically 0.3 nA for the current set point and 0.8 V for the bias
voltage, $V$. This implies that we probed the unoccupied
electronic DOS of PPMO. Fig.~\ref{Iz}(a) shows the dependence of
tunneling current $I$ on relative tip-sample distance $z$ on a
semi-logarithmic plot. The exponential nature of $I(z)$ confirms
an excellent vacuum tunnel barrier (effective work function $\phi
\sim$ 1.5 eV). Topography (50 $\times$ 50 nm$^2$) is presented in
inset (b). Terraces with unit cell height ($\sim$0.4 nm) steps
[Fig.~\ref{Iz}, inset (c)] indicate $\langle 100 \rangle$ surface
of the pseudocubic perovskite crystal.

\begin{figure}[tb]
\centering \includegraphics[width=8.0cm,clip]{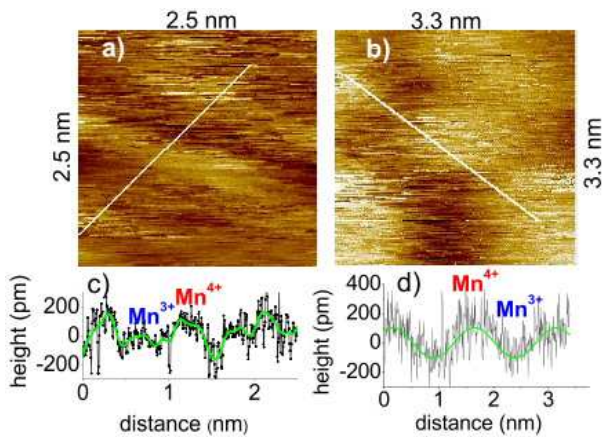}
\caption{(a), (b) Topography of different areas in the insulating
regime ($T =$ 300 K). (c) and (d) show corresponding intensity
profiles along the white lines drawn in (a) and (b),
respectively. The bright and dark spots in the image are
associated with Mn$^{4+}$ and Mn$^{3+}$ ions, respectively.}
\label{polar}
\end{figure}

High resolution STM images taken in the insulating regime (at 300
K) on the terraces indicate bright and dark regions forming
stripe-like features spread over a length scale of 0.4--0.5 nm, as
seen in Fig.~\ref{polar}(a), (b). While probing the unoccupied
electronic DOS the doped holes localized on Mn$^{4+}$ sites
appear as bright spots in the STM image, whereas electron
tunneling from a conventional metallic tip into a polaronic state
(e.g. electrons localized on Mn$^{3+}$ ions) is difficult and
produces dark spots \cite{ma}. However, these contrasts were seen
only occasionally, an observation similar to what was reported in
the case of layered manganite La$_{2-2x}$Sr$_{1+2x}$Mn$_2$O$_7$
\cite{ron}. This suggests the short-range stripe-like order of
Mn$^{3+}$ and Mn$^{4+}$ ions. The extent of these features
[Fig.~\ref{polar}(c) and (d)] is slightly larger than the typical
atomic distance of $\sim$0.39 nm in the cubic perovskite cell and
comparable with charge ordered stripes observed in high
resolution lattice images of La$_{0.33}$Ca$_{0.67}$MnO$_3$
\cite{mor}. Short-range polaron correlation and CE-type of charge
ordering was observed in manganites using diffused x-ray and
neutron scattering \cite{vas,nel}. Recent STM studies \cite{ma}
probing simultaneously the occupied and unoccupied states of
(La$_{5/8-x}$Pr$_x$)Ca$_{3/8}$MnO$_3$ thin films also showed
short-range CE-type charge ordered clusters in the pm state.

To map the surface electronic state, we carried out thousands of
STS measurements at different locations on the sample
\begin{figure}[tb]
\centering \includegraphics[width=7.85cm,clip]{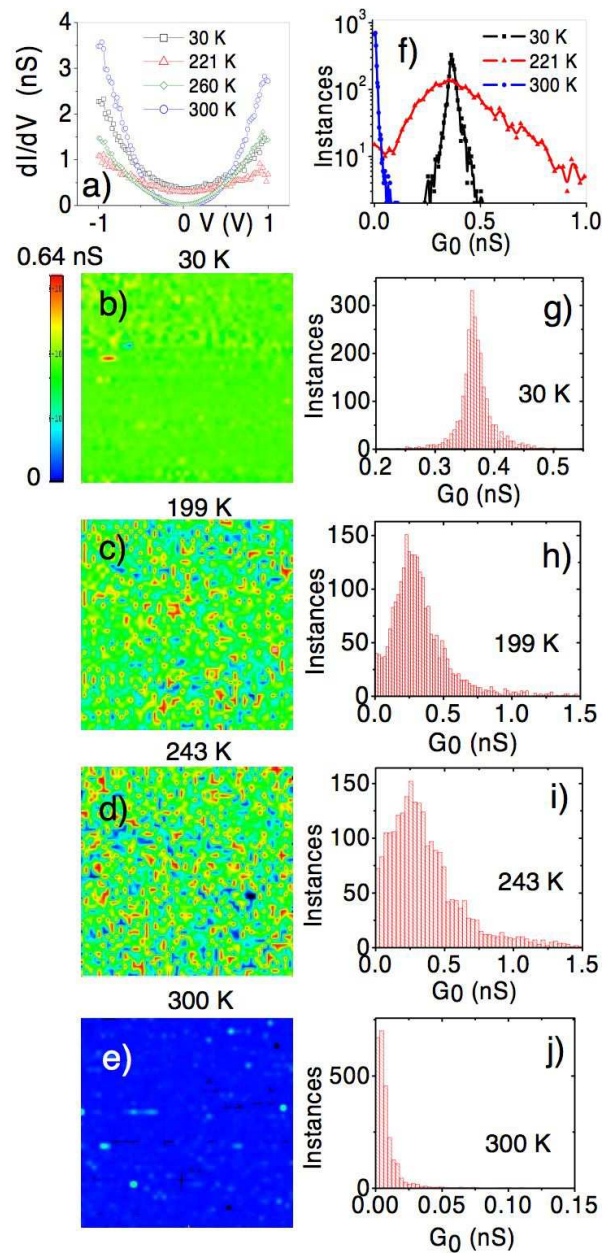}
\caption{(a) $G$-$V$ curves averaged over an area of 50 $\times$
50 nm$^2$ at 30, 221, 260 and 300 K indicating metallic
($T<T_{MI}\approx$ 255 K) and semi-conducting behavior
($T>T_{MI}$). Panels (b)--(e) present conductance maps (50
$\times$ 50 nm$^2$) taken at 30, 199, 243 and 300 K through $T_C$
and $T_{MI}$. The color scale shown left of panel (b) encodes
$G_0$. (g)--(j) Histograms of $G_0$ at the same $T$ as presented
in (b)--(e). (f) Histograms for the three different $T$ regimes
compared on a semi-logarithmic scale.} \label{spectro}
\end{figure}
surface spanning the $T$ range from 28 -- 300 K. Typically, a
surface area of 50 $\times$ 50 nm$^2$ with a lateral resolution
of 1 nm (2500 pixels) was investigated. Tunneling current and
differential conductance, $G = dI/dV$, were measured
simultaneously while ramping $V$ from $-1$ to +1 V. An average of
2500 $G$-$V$ curves taken at representative $T =$ 30, 221, 260 and
300 K are shown in Fig.~\ref{spectro}(a). At 30 K, the $G$-$V$
curve is metal-like with a finite value of $G_0$ signifying a
finite DOS at the Fermi energy. In contrast, at 300 K, i.e. $T >
T_{MI} \approx$ 255 K, the $G$-$V$ curve around $V =$ 0 is
typical of a semiconducting gap. In the pm metallic state at 221
K, the value of $G_0$ is only slightly reduced compared to $G_0
(T=30$ K$)$, thus showing an expected $T$ dependence of the
conductance.

For quantifying the STS results and mapping the homogeneity of the
DOS laterally as well as its temperature evolution, we plot $G_0$
as conductance maps at selected $T$. In
Fig.~\ref{spectro}(b)--(e), the {\em local} $G_0$ is presented
color-coded (with a color scale covering 0 $\le G_0 \le$ 0.64 nS)
at $T=$ 30, 199, 243 and 300 K, respectively. Corresponding
distributions for the frequency of the observed $G_0$ values are
shown in the histograms Fig.~\ref{spectro}(g)--(j). A sharp
distribution of $G_0$ at 30 K confirms a homogeneous electronic
phase at low temperature. Similarly, the conductance map at 300 K
(in the semiconducting region) is also homogeneous
[Fig.~\ref{spectro}(e)], with most of the values of $G_0$ very
close to zero [Fig.~\ref{spectro}(j)].

On the other hand, as $T$ is raised through $T_C$ and approaches
$T_{MI} \approx$ 255 K inhomogeneities start to develop at a
length scale of 2 -- 3 nm, as seen in Fig.~\ref{spectro}(c) and
(d). The peak in the histograms [panels (h), (i)] shifts to lower
conductance values as $T$ is increased and, importantly, an
increasing weight at $G_0 \rightarrow$ 0 is observed. A bimodal
distribution of $G_0$ at $T =$ 221 K is clearly visible in
Fig.~\ref{spectro}(f), with two maxima in $G_0$ frequency located
at similar $G_0$ values as for low and high $T$, respectively. The
distributions near $T_{MI}$ [cf. Fig.~\ref{spectro}(f)] are
significantly broadened compared to both, low $T$ (30 K) and high
$T =$ 300 K $> T_{MI}$. The sharp distribution at $T =$ 300 K
clearly indicates that these broad distributions of $G_0$ at
intermediate $T$ reflect a sample property rather than an
instrumental influence.

The $T$ dependences observed in STS arise not only from the Fermi
function, but also the sample's DOS itself is $T$ dependent. This
change of electronic properties can be explained by the release
of lattice distortions around $T_{MI}$, when the immobilized
polaronic carriers become successively mobile producing
inhomogeneous spatial conductance distributions and electronic
transport through percolating metallic regions. Thus, the
increasing weight at $G_0 \rightarrow$ 0 while retaining a peak at
$G_0 \sim$ 0.3 nS provides a direct observation of nanometer-scale
PS in PPMO single crystals. Interestingly, this PS appears to be
restricted to the transition region $T_C \lesssim T < T_{MI}$ in
this compound. The drastic change in $G_0$ and its distribution
with $T$ at around the bulk $T_{MI}$ indicates that our STS
results are not mere surface effects.

Our results are distinct from previous experimental results where
PS is seen on a micrometer scale and persisted well within the
metallic regime. It remains an open question, whether the
particular properties of PPMO with a metallic pm state in the
region $T_C < T_{MI}$ are responsible for the clear observation
of this nanometer-scale PS phenomenon and whether the result can
be generalized to other mixed-valence manganites (as pointed out
before the electrical transport in PPMO occurs via percolation of
nanometer-scale metallic clusters for $T_C < T < T_{MI}$).
Further, the specific pattern of electronic inhomogeneitiy in the
local surface DOS is certainly affected by disorder, induced by
random chemical substitutions and/or surface effects. In
addition, disorder effects due to size differences between
$A$-site Pr$^{3+}$ and Pb$^{2+}$ ions may play a role. However,
the observed nanometer-scale PS is not a simple and fixed result
of static chemical disorder, as can be inferred from the
homogeneity of the electronic properties deep in the metallic
state (low $T$) as well as in the insulating one (300 K). Hence,
in order to resolve the relevance of disorder effects on PS and
the associated length scale, similar spatially resolved STS
studies on different manganites are called for.

In summary, our high resolution STM images provide direct
evidence for the high temperature polaronic state in perovskite
manganite. Polarons are confined to one lattice cell. Stripe-like
features seen occasionally in these images suggest a short-range
ordering of these polarons in the form of a lattice. Spatially
resolved STS images show nanometer-scale phase separation in the
paramagnetic metallic state. However, the homogeneous low-$T$ as
well as high temperature STS images confirm that this phase
separation is limited only to the temperatures close to the
metal-insulator transition suggesting that it is related to the
non-coincidence of $T_C$ and $T_{MI}$.

We are grateful to U.~K. R\"o\ss ler, Ch. Renner and G. Aeppli for
valuable discussions. We acknowledge financial support by the
DFG, grant WI 1324/1-1, and the European Commission through
CoMePhS 517039.

\end{document}